\documentclass[letterpaper]{article}
\usepackage{spconf,amsmath,graphicx}
\usepackage{booktabs,xcolor,array,longtable}
\usepackage{flafter} %
\usepackage[hidelinks]{hyperref}
\makeatletter
\g@addto@macro\UrlBreaks{\do\-}
\makeatother

\title{Extending Music Annotation Schemas:\\Zero-Shot Prediction or Few-Shot Adaptation?}
\name{Christos Plachouras  \thanks{Christos Plachouras is supported by UKRI grant number EP/S022694/1.} \qquad Emmanouil Benetos \qquad Johan Pauwels}
\address{School of Electronic Engineering and Computer Science, 
Queen Mary University of London, UK\\
\{c.plachouras, emmanouil.benetos, j.pauwels\}@qmul.ac.uk}

\begin{document}
\ninept
\maketitle

\begin{abstract}
Automatic music annotation is typically tackled under the assumption of a fixed annotation schema. In practice, commercial music catalogs often need to accommodate new musical attributes as needs evolve. Given that expert music annotation is expensive, it is not evident which methodological approach is most effective at accommodating new attributes and backfilling existing tracks; audio-language models promise zero-shot prediction, but at what annotation budget does supervised adaptation become more compelling?

We propose a benchmark based on the MGPHot popular music annotation dataset for simulating music schema extension across different annotation budgets. We investigate zero-shot prediction with audio-language models, learning new attributes from pretrained representations, and adapting models trained on existing annotations. Our results suggest that supervised adaptation is more effective than zero-shot prediction even with small annotation budgets, while frozen representation reuse remains the most effective approach for modest budgets without the tuning required by deeper adaptation.

\end{abstract}


\begin{figure}[t]
  \centering
  \includegraphics[width=\columnwidth]{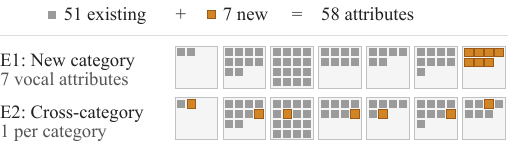}
  \caption{Two schema extensions on MGPHot. Boxes denote categories. Both extensions add 7 attributes to an existing set of 51.}
  \label{fig:schema-extension}
\end{figure}

\section{Introduction}
\label{sec:introduction}

Music tagging involves assigning descriptive labels to music tracks to describe their musical attributes, such as the genre, mood, and instrumentation. It is a long-standing task in music information retrieval (MIR)~\cite{levyMusicInformationRetrieval2009,turnbullSemanticAnnotationRetrieval2008}, providing a compact, semantic representation for organizing, searching, and recommending music~\cite{firan_benefit_2007,wang_tag-based_2018}. Automatic prediction learned from a manually labelled subset has made tagging scalable to large collections~\cite{won2020eval,landau2023mgp2}.

However, expert annotation is expensive, taking several minutes per track and requiring rigorous protocol adherence and quality control~\cite{oramas2025mgphot}.
The annotation schema---the attributes, their definitions, and their allowed values---is often updated to accommodate emerging musical styles, new sounds, evolving user needs, and changing service requirements~\cite{landau2023mgp2,ddexMediaEnrichment}. In this paper, we study the resulting problem: given a music catalogue with existing annotations, how can we extend the schema to include new attributes and predict their values for the entire collection?

One way to extend the schema is to adapt a model trained on the existing attributes using a small set of new annotations. However, limited examples can lead to overfitting, while updating the model can cause forgetting~\cite{li_learning_2018,gonzalez-barrachinaContinualLearningMusic2024}. A lighter alternative is to freeze the audio encoder and train only new prediction heads. This preserves the existing representations, but they may lack information needed for the new attributes. An encoder pretrained on larger audio collections may provide more general features instead~\cite{li2024mert}. Still, both approaches require new annotations. Audio-language models offer a compelling zero-shot alternative, using their extensive audio-language pretraining to predict from attribute descriptions without new annotations or model adaptation~\cite{wuLargeScaleContrastiveLanguageAudio2023,ghosh2025musicflamingo}. How accurately can these models reproduce the required ratings, and at what annotation budget do supervised approaches outperform them?

We compare these approaches using the expert attribute ratings from the MGPHot dataset~\cite{oramas2025mgphot}, under shared experimental conditions and annotation budgets of 8, 32, and 128 support tracks. To our knowledge, this is the first benchmark to study music schema extension in an industry-motivated setting across contrasting method families. Our contributions are:
\begin{list}{\textbullet}{%
    \setlength{\leftmargin}{1em}%
    \setlength{\labelsep}{0.4em}%
    \setlength{\itemsep}{0pt}%
    \setlength{\parsep}{0pt}%
    \setlength{\topsep}{2pt}%
    \setlength{\partopsep}{0pt}}
    \item A reusable benchmark protocol for extending music annotation schemas with limited annotations for new attributes.\footnote{Available at \href{https://github.com/chrispla/music-schema-extension}{https://github.com/chrispla/music-schema-extension}}
    \item A systematic comparison of zero-shot audio-language prediction, frozen representation reuse, and adaptation of models trained on existing attributes.
    \item An analysis of annotation-budget effects, retention of existing attributes under adaptation, and sensitivity to possible pretraining exposure to test recordings.
\end{list}

\section{Background}
\label{sec:background}

Automatic music tagging includes both binary label prediction~\cite{won2020eval} and regression of perceptual ratings~\cite{aljanaki2018midlevel}.
Over the past decade, the MIR community has shown interest in large-scale supervised and self-supervised learning to produce general-purpose audio representations that transfer to music tagging~\cite{hershey2017cnn,li2024mert}.
Still, annotation quality remains a concern, with missing and incorrect tags~\cite{choi2018noisylabels}.
More recently, MGPHot~\cite{oramas2025mgphot} introduced expert ratings for 58 musical attributes, with defined scales and quality checks that support evaluation of attribute strength beyond binary presence.
A subsequent benchmark~\cite{ramoneda2025benchmarkingmusicautotaggingmgphot} compares frozen representations from several recent models on tagging and regression.
We build on this setting to study schema extension with limited annotations for new attributes.

Extending an annotation schema can be framed as a problem tackled by different approaches in machine learning.
Continual learning methods focus on updating models to learn new tasks while avoiding performance degradation on previous ones, a phenomenon known as catastrophic forgetting~\cite{kirkpatrick_overcoming_2017}. They have previously been explored in a limited setting for music classification~\cite{gonzalez-barrachinaContinualLearningMusic2024} and more generally in audio classification~\cite{wangFewShotContinualLearning2021,mulimani2025closerlook}.
Metric learning approaches instead focus on learning a representation space where new classes can be defined from a few examples, without retraining the model for each extension. Music applications have included instrument recognition~\cite{garciaLeveragingHierarchicalStructures2021} and multi-label tagging~\cite{papaioannouLCProtonetsMultiLabelFewShot2025}, though no extension to regression tasks has been proposed.

More recently, attention has shifted towards joint audio-text embeddings~\cite{choiZeroshotLearningAudiobased2019a,manco2022muscall,huang2022mulan,wuLargeScaleContrastiveLanguageAudio2023} or generative models such as Music Flamingo~\cite{ghosh2025musicflamingo}, which can leverage natural language descriptions to predict attributes without retraining.
While these approaches exhibit strong performance on audio-language benchmarks, it is still understudied how well they perform under stricter experimental conditions, including for numerical ratings, and in comparison to other approaches. CMI-Bench~\cite{ma2025cmibench} demonstrates their underwhelming performance on a range of MIR tasks, with more positive results on high-level musical attribute classification.
Further, there are concerns about data contamination in the training of these models, which may lead to overestimation of their performance on downstream tasks~\cite{wu2025clamp3}.
This motivates our comparison under common annotation budgets and evaluation conditions, including a check for possible audio pretraining overlap.

\section{Experimental setup}
\label{sec:experimental_setup}
\subsection{Data and schemas}
We select MGPHot for its expert musicological annotations and graded attribute scores. It provides continuous ratings in $[0,1]$ for 58 attributes of 21,320 tracks across rhythm, instrumentation, sonority, harmony, compositional focus, vocals, and lyrics~\cite{oramas2025mgphot}. We compare two schema extensions with lower (E1) and higher (E2) expected semantic relatedness to existing attributes, using category membership as a proxy. Each adds seven target attributes to 51 base attributes (Fig.~\ref{fig:schema-extension}). E1 holds out all seven vocal attributes as targets: register, timbre, breathiness, smoothness, grittiness, nasality, and accompaniment. E2 adds one attribute per category: vocal breathiness, harmonic sophistication, syncopation, electric guitar distortion, explicit lyrics, synthetic sonority, and focus on riffs.

To measure performance differences across different annotation budgets, we create support sets $\mathcal{S}_K$, where $K\in\{8,32,128\}$ specifies the number of tracks with annotations for all seven target attributes. MGPHot's canonical artist-disjoint splits contain 12,800 training tracks and 4,286 test tracks. We sample support tracks from the training split in eight random draws to capture variation across support sets, and evaluate on test recordings. We use macro RMSE over the seven target attributes and report the mean and sample standard deviation across draws.

\subsection{Validation}
Given the small support sets, holding out tracks for validation would further reduce the available training data while providing too few examples for reliable model selection. Instead, we consider the set of 45 attributes shared by the E1 and E2 base schemas, and use it and its validation set to select hyperparameters and other method settings, as described further in Sec.~\ref{sec:methods}. We then apply the selected setup to the target attributes.

For deeper model adaptation, we also need to select hyperparameters for learning new attributes while retaining performance on existing ones. We therefore simulate two extensions by training base models on 38 of these attributes and adapting them to seven held-out attributes. One simulation holds out a category, while the other selects attributes across categories, mirroring the two evaluation extensions.

\begin{figure*}[t]
  \centering
  \includegraphics[width=\textwidth]{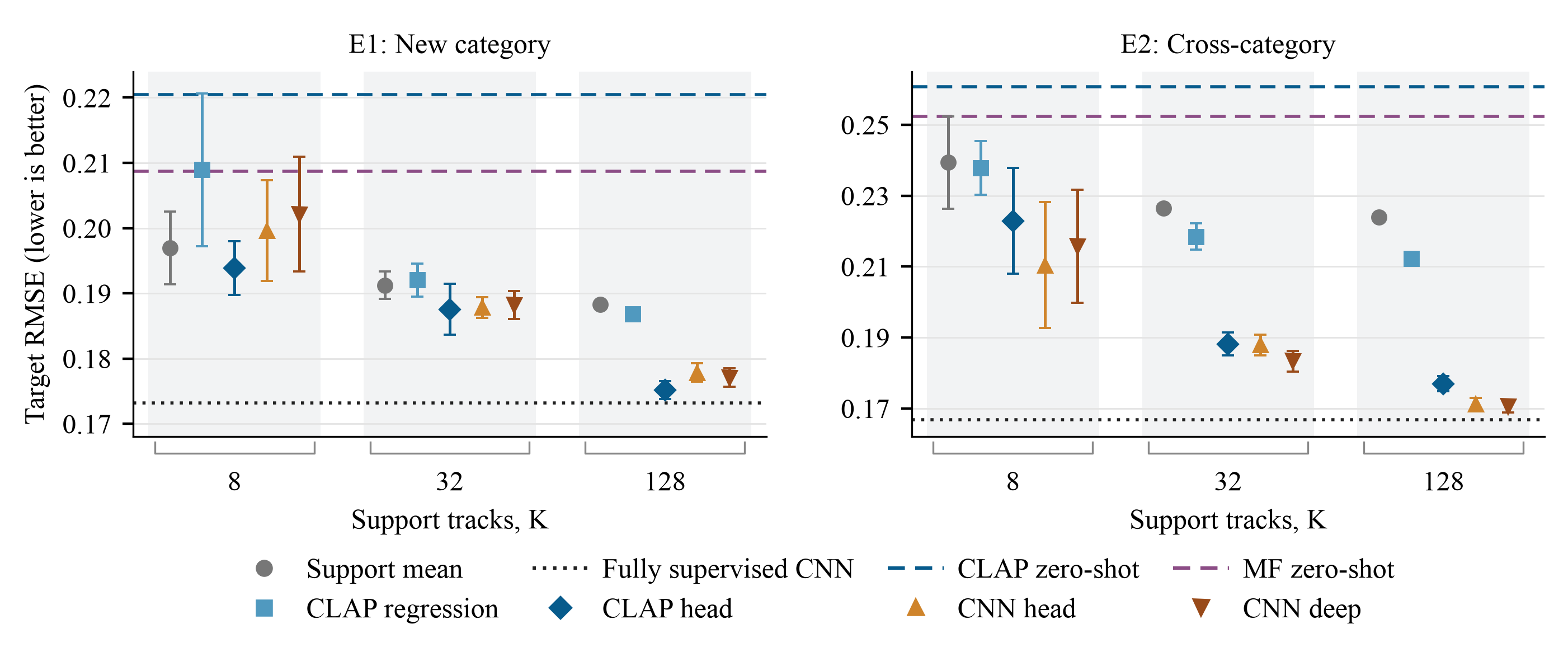}
  \caption{Prediction error on the seven target attributes for each extension. Points show mean RMSE with $\pm1$ sample standard deviation across eight draws. Horizontal lines show zero-shot methods and the fully supervised CNN, which is outside the support budget. Lower is better.}
  \label{fig:benchmark-results}
\end{figure*}

\subsection{Methods}
\label{sec:methods}
In this section, we describe two zero-shot approaches we use as well as an extension of one to leverage support tracks, the representation reuse approach using two models, and the deeper adaptation approach using one model. These are compared to two baselines using no or all available audio to contextualize the results.
\subsubsection{CLAP zero-shot} CLAP learns a shared audio-text representation, allowing us to compute a similarity between test recordings and attribute descriptions~\cite{wuLargeScaleContrastiveLanguageAudio2023}. However, as indicated by studies with its visual counterpart CLIP, these single-prompt similarities can have poor correlation with human ratings ~\cite{wang2023clipiqa}. Thus, similar to CLIP-IQA~\cite{wang2023clipiqa}, we use each attribute's definition to create text anchors describing its rating range, using either two endpoints ($0,1$) or five levels ($0,0.25,0.5,0.75,1$). For example, for `tempo', five anchors describe ``very slow'', ``slow'', ``moderate'', ``fast'', and ``very fast'' at values $0,0.25,0.5,0.75,1$. A softmax over audio-text cosine similarities weights the anchor ratings for each window, and we average these predictions across the recording:
\begin{equation}
\hat y(x)=\frac{1}{W_x}\sum_{w=1}^{W_x}\sum_{k=1}^{A}v_k\,\operatorname{softmax}(s_w/T)_k,
\label{eq:clap}
\end{equation}
where $W_x$ is the window count, $s_w$ the similarity vector, and $v_k$ the rating of anchor $k$ among $A$ anchors. Lower temperature $T$ gives more weight to the most similar anchors. The weighted average allows continuous predictions between anchor values.

Since CLAP struggles with negation modeling~\cite{vasilakis2026negation}, we also create alternative affirmative anchor descriptions that avoid negation but occasionally sacrifice some precision (e.g. for `backbeat', turning \textit{`No backbeat emphasis on beats two and four'} into \textit{`Rhythmic emphasis on the downbeat'}, which need not be true). We compare four checkpoints and alternative anchor wording and prompt styles. Validation selects the music-speech-AudioSet checkpoint, five anchors that include negatives, prompts of the form ``A music recording with \ldots'', and $T=0.07$. Five anchors consistently achieved lower validation RMSE across configurations. Other settings, including the checkpoint, had minimal differences after temperature optimization.

\subsubsection{Music Flamingo zero-shot} Music Flamingo (MF) can receive audio of up to 10 minutes and a text instruction to produce an output~\cite{ghosh2025musicflamingo}. We try two approaches. First, we instruct it to produce a rating between 0 and 1 for the attribute, given its definition, optionally adding three text anchors at $0$, $0.5$, and $1$ to clarify the range. Second, inspired by probability-based visual scoring in Q-Bench and Q-Align~\cite{wu2024qbench,wu2024qalign}, we use two or five anchors from the original CLAP descriptions. We assign a letter to each anchor option and ask MF to return the letter of the best match. We extract the next-token logits for the option letters and apply a softmax at $T=1$, using the probabilities to weight the anchor values. Through validation, we select the five-anchor option-based formulation, which slightly improves RMSE over definition-only numeric prediction. Adding anchors worsens numeric prediction; using two options also gives higher error than the selected five-option formulation.

\subsubsection{CLAP similarity regression} We also test whether support tracks can improve the mapping from CLAP audio-text similarities to ratings. Keeping CLAP frozen, we average similarities across windows and fit a linear regression with an intercept using $\mathcal{S}_K$. We compare three inputs: the high-minus-low endpoint difference, the two endpoint similarities separately, and all five anchor similarities. Selection uses validation RMSE averaged across shared base attributes, support draws, and budgets. Validation selects the endpoint difference, using the music-AudioSet checkpoint and prompts of the form ``This music has \ldots''. Across all 24 matched configurations, five inputs improve validation error at 128 support tracks but worsen it at eight and 32.

\subsubsection{Frozen audio heads} As a more conventional few-shot learning approach, we learn ratings from frozen audio features by fitting one linear regression head per attribute. Given the network is frozen apart from the independent regression for each attribute, we inherently avoid any forgetting considerations during adaptation with this setup. We use the 512-dimensional audio embeddings of the pretrained CLAP encoders from the four checkpoints, and the 512-dimensional features from the shared dense layer of a ShortChunkCNN with seven convolutional blocks~\cite{won2020eval}. We pretrain the CNN from random initialization to predict the 51 base attributes in $\mathcal{B}_{\mathrm{E1}}$ or $\mathcal{B}_{\mathrm{E2}}$, separately for each extension. We use the base validation set RMSE to select a learning rate of $3\times10^{-4}$ and the best checkpoint. During pretraining for the CNN and head training for the CNN and CLAP, we randomly sample a window from each recording in each epoch, using their native window lengths of 3.69 and 10 seconds, respectively. We L2-normalize CLAP embeddings, consistent with its cosine-similarity objective.

To limit overfitting with few support tracks, we use ridge regularization, which penalizes large weights. We fit each head on $\mathcal{S}_K$ by minimizing
\begin{equation}
\min_{\mathbf w,b}\;\frac{1}{K}\sum_{i=1}^{K}(y_i-\mathbf w^{\top}\mathbf z_i-b)^2+\lambda\|\mathbf w\|_2^2,
\label{eq:ridge}
\end{equation}
where $\mathbf z_i$ and $y_i$ are the averaged audio features and rating of track $i$. The intercept $b$ is unpenalized. We select the regularization strength $\lambda$ separately for each support budget using validation RMSE averaged across the shared base attributes and support draws. Averaging these scores across budgets selects the 630k-AudioSet CLAP checkpoint. The selected ridge penalties are $0.1$, $0.01$, and $0.01$ for CLAP at $K=8,32,128$, respectively. The CNN heads use $1$, $1$, and $0.1$ for both extensions.

\subsubsection{CNN adaptation} Finally, we test whether it is beneficial to update the CNN more deeply to predict new attributes, given the limited available adaptation data. We start from the base CNN and continue training on the support tracks, updating either the last 1, last 4, or all 7 convolutional blocks, together with the shared dense layer and target heads. The base output heads remain fixed. During adaptation, the network also learns from the base-attribute labels of the support tracks, which helps alleviate catastrophic forgetting. We also try explicit experience replay~\cite{rolnick2019experience}, by including 256 training tracks with target attributes hidden during each adaptation.

Adaptation minimizes squared error, averaged separately over new and base attributes and then weighted equally, to encourage learning the smaller target attribute set. In addition to depth and replay, we vary the learning rate and number of updates. The target heads use ten times the learning rate of the shared layers. Selection balances validation RMSE on base and simulated target attributes in the ratio $51:7$, reflecting the final extensions. We average this score across both simulations and their support draws, selecting settings separately for each support budget. The selected configuration updates only the final convolutional block, shared dense layer, and target heads, with 256 replay tracks and a shared-layer learning rate of $10^{-5}$. It uses 25, 25, and 200 updates for $K=8,32,128$, respectively. We measure forgetting as base-attribute RMSE after adaptation minus its value before adaptation, with positive values indicating reduced accuracy.

\subsubsection{Baselines and reference} The support-mean baseline uses the mean attribute value in $\mathcal{S}_K$ as its prediction for every test track. It uses no audio. We also train the same CNN architecture from random initialization to predict all 58 attributes jointly, using their labels across the full training set. This provides a reference for the gap to performance using the full training set for the given method. We reuse the learning rate selected for CNN base training and select its training epoch by validation RMSE over all 58 attributes.

\section{Results}
\label{sec:results}

\begin{figure}[t]
  \centering
  \includegraphics[width=\columnwidth]{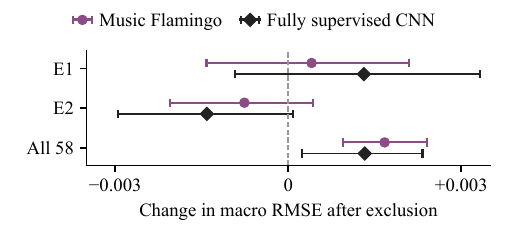}
  \caption{Macro RMSE change after excluding possible audio pretraining matches, for E1/E2 targets and all 58 attributes. Both models use the same retained tracks. Bars are 95\% paired recording-bootstrap intervals. Positive values mean higher error.}
  \label{fig:audio-overlap-results}
\end{figure}

Figure~\ref{fig:benchmark-results} compares prediction error across the two schema extensions and three annotation budgets. MF consistently outperforms zero-shot CLAP, but even the mean ratings of support tracks achieve lower average error in both extensions, showing that some of the benefit of annotation comes from estimating the rating distribution of the new attributes. Fitting a regression to CLAP similarities reduces error as more support tracks become available, though its advantage over the support mean is limited, particularly for E1.

Fitting heads to frozen audio features generally improves predictions further, with both encoders outperforming CLAP similarity regression at every budget, suggesting that CLAP’s audio representations contain predictive information that the tested text-similarity methods do not fully exploit. The encoders' relative performance differs between extensions. CLAP gives lower error for the new-category extension (E1), with its largest advantage over the CNN at eight support tracks. Its broader pretraining may provide features that better bridge the semantic gap to attributes outside the existing categories, particularly when few annotations are available. Conversely, the CNN performs better in the cross-category extension (E2), where training on related base attributes may provide features more suited to the new targets. Although these patterns suggest that the semantic relatedness of new attributes affects adaptation accuracy, further controlled comparisons are needed to establish the cause. At 128 tracks, both encoders approach the fully supervised CNN reference.

Updating the CNN more deeply gives small, mixed changes compared to fitting frozen heads. Both eight-track comparisons worsen, while the largest mean improvement is 0.0046 RMSE for E2 at 32 tracks. This suggests that updating the representation may be less effective when very few examples are available, although additional support does not consistently lead to an advantage over frozen heads. Changes in base-attribute RMSE range from $-0.000186$ to $+0.001028$, indicating that existing predictions are largely retained. Under the tested settings, frozen heads therefore offer competitive target predictions while preserving the base model.

\label{sec:pretraining}

Audio-language models may have encountered some test recordings during pretraining. We examine this possibility using the documented training sources for MF and CLAP. For CLAP, we use the 630k-AudioSet checkpoint because its training-record lists are publicly available, unlike those of the other checkpoints.

To assess audio overlap, we identify exact video-ID matches or exact agreement in normalized artist and song title. The latter indicates possible exposure to the song, though not necessarily the same recording. We remove recordings matched to either model's training sources and recompute error on the same retained subset. Only three recordings match the documented sources for this CLAP checkpoint, so we focus on MF. Figure~\ref{fig:audio-overlap-results} shows the changes after this shared exclusion, which leaves 2,480 recordings; MF's target RMSE changes by only $+0.0004$ for E1 and $-0.0008$ for E2. Across all 58 attributes, RMSE increases by $0.0017$ for MF and $0.0013$ for the fully supervised CNN. However, removing tracks can also change the difficulty of the test set. We therefore compare MF's error change with that of the fully supervised CNN on the same retained tracks, using paired recording bootstrap intervals. The 95\% intervals for the difference include zero in both extensions, providing no clear evidence of a differential effect. The same holds across all attributes, where the difference in RMSE changes is $+0.0003$ (95\% interval: $[-0.0005,\,+0.0012]$).

\section{Conclusion}
\label{sec:conclusion}

We introduced a publicly available benchmark for extending music annotation schemas and compared zero-shot prediction with supervised adaptation under common annotation budgets. On the two MGPHot extensions, even a simple baseline outperformed CLAP and Music Flamingo. With a few annotations, fitting heads to frozen audio features gives further improvements, with a CNN trained on existing attributes remaining competitive with pretrained CLAP representations. Updating the CNN more deeply offers smaller, less consistent benefits. These findings support reusing existing audio representations with limited new annotation as a practical approach to schema extension. They also call for further research into improving audio-language models for non-conversational tasks, like rating prediction. We hope the benchmark provides a basis for testing and improving this trade-off between annotation cost and prediction accuracy in music.

\section{Compliance with Ethical Standards}
This study analyses existing music recordings and annotations from MGPHot~\cite{oramas2025mgphot}. No ethical approval was required.

\section{Acknowledgements}
The authors declare no conflicts of interest.

\bibliographystyle{IEEEbib-compact}
\bibliography{refs}

\end{document}